# Authenticated Byzantine Gossip Protocol

## Egor Zuev

zyev.egor@gmail.com

## Abstract

ABGP refers to Authenticated Byzantine Gossip Protocol. The ABGP is a partial-synchronous, weak consistent, BFT based consensus algorithm. The algorithm implements the gossip protocol, but with BFT features inside (like multisig record approval). The algorithm has been developed as an alternative to classic private ledger solutions, like Hyperledger.

**Keywords:** consensus algorithms, distributed systems, RSM, DLT.

## 1. Introduction

ABGP is a partial-synchronous, weak consistent BFT algorithm. The algorithm implies that all nodes in network are known. To make sure, that there is no malicious node connected to network, ABGP use authentication approach. This means, that all proposed state mutations should be signed by nodes who propose it, or accept (until the final mutlisignature will be built). The replication between nodes happens by standard gossip protocol in one step: one node request for new changes from another, and another node sends back either new changes, or an empty array, if there are no new changes. On new record replication, the acceptor node, validate this record's signature, in order to make sure that it has been signed by known nodes. The algorithm offers both liveness and safety for at most $N = 2f + 1$ with quorum of $Q = f + 1$, where N – the number of nodes in cluster and $f$ – is the number of faulty nodes in case of N-to-N connections. However, in case of M-of-N connections, the quorum may be much smaller.

## 2. System model

The system model represents a partially synchronous system, where nodes are connected by the network. We assume that the network may duplicate, fail to deliver, or deliver with delays messages. We also assume, that every node may become failed. By failed we mean malicious behavior, or just out-of-order.

The system is partially synchronous, which means that we expect delays / duplicates / message lost in the system. The communication between node is bi-directional, which means, if node A sends message to node B, then node B should send message to node A as a reply (like an HTTP request).

The system requires to be aware of all other nodes in the system. It means, that each node should have a bootstrapped list of all other nodes in the network. This bootstrap list should include address of node (which is optional in case of M-of-N connections) and public keys of all other nodes.

The algorithm uses ECC cryptography for signing purpose. Once enough nodes sign the hash of the record, the last node, which signed the hash – creates a multisignature. This is a part of "authentication mechanism".

The term "authentication" means, that all mutation actions in algorithm should be signed and validated by its members (i.e. nodes). While "sign" action – means signing the record, the validation – means signature / multisignature validation.

The replication process inherits the gossip protocol properties: node choose randomly another peer in network and ask recent updates.

The algorithm guarantees **safety and liveness** as long as $N \geq 2f + 1$ for N-of-N connections. The safety property means that all healthy nodes agree on the certain change, sign it and apply to the state. The liveness means that even if $f$ nodes fail, the network will still be operatable.

## 3. Initial node state

Before the boot phase, each node should have the following properties (or settings):

| Property | Description | Data type | Example |
|---|---|---|---|
| Private key | generated with secp256k1 standard | String | "edca08af903bbd7afd356bb2d6fdf54901fad8770c734416456d382012b0608b" |
| Public key | generated with secp256k1 standard (from private key) | String | "02b68e80e752f13906179688171d6a0345a6c6edd20b52f24ea489275668fa1eac" |
| Min gossip interval | Minimum gossip interval (ms) | Number | 150 |
| Max gossip interval | maximum gossip interval (ms) | Number | 300 |

| Proof expiration | How long the proof is valid (ms) | Number | 300 |
|---|---|---|---|
| Nodes | An array of nodes (peers) addresses (should include network address + public key of the peer) | [string] | [<br>"tcp://127.0.0.1:2000/ 036032cc7790000ac98fc2dbc200c486b073e74e8e4ff5fba15beb01c153f4c458"<br>] |

Consensus parameters

After the boot phase, the gossip protocol starts and local API (for appending new records) becomes available.

## 4. Append process

### 4.1 Local append

The append process looks like that:

1) Each new record should have key, value and version fields
2) On append, the algorithm should create hash of record: $hash = sha256(key, value, version)$. This hash brings uniqueness to the record
3) Then algorithm should create partial signature as follow: $\text{partialSignature} = (\text{privateKey} * \text{hash}) \mod N$, where $N$ is curve parameter.
4) Then algorithm add timestamp and timestamp index to the record. The timestamp – is a timestamp when record is created. The timestamp index is used when concurrency is possible, and two or more records can be created at the same time. In case this happens, all records with the same time have their own index (like 0, 1, 2).
5) Then this record, alongside with hash and signature can be stored locally (for instance in database)
6) This record is called intermediate

```
RecordModel {
hash: '025fe4bbee09ea3e9933ee43982fa25d4a83180ae11f336c6ee75a6dc3c58d11',
key: '52f93649ed8ee14',
```

```
  value: '52f93649ed8ee14',
  version: 1,
  signaturesMap: {
    '0270633bb456e6b717f309caa6429b892717cec7bae1f3a6b14ce6b0923ae83b58' => 'b0f54870c494801b8cb5b0c3784bc6ec895ae240ab60b9cbde79d242c2c07796'
  },
  timestamp: 1653722582280,
  timestampIndex: 0,
  signatureType: INTERMEDIATE,
  publicKeys: [
    '0270633bb456e6b717f309caa6429b892717cec7bae1f3a6b14ce6b0923ae83b58'
  ]
}
```

Example of intermediate record

### 4.2 Append from another node

When one node receives new records from another (for instance node A obtained records from node B) during replication process, the append rules vary:

1) The algorithm should validate the record
2) Then algorithm should check, do this node already has this record (it can be done by finding the record by hash).
   2.1) If record exist then:
      2.1.1) In case received record is multisig and local record is intermediate – then algorithm should replace local intermediate record with received multisig and update the root.
      2.1.2) In case local and received records are multisig, then the highest multisig is chosen (the algorithm compares 2 signatures by value) and stored in local record.
      2.1.3) In case local record is multisig and received one is intermediate – then algorithm ignore this record (i.e. doesn't apply)
      2.1.4) In case local and received records are intermediate – then algorithm just take signatures from received record (which are not present on local record) and append them to local record.
   2.2) if record doesn't exist:
      2.2.1) then algorithm should sign the hash of the received record (like was in local append described above), add it to this record and store
3) Then the algorithm should check if there are enough signatures for multisig (this number is defined by quorum size).
   3.1) if yes then:

3.1.1) algorithm build multisig: $signature = \sum partialSignature_i \bmod N$

3.1.2) algorithm build shared public key: $sharedPublicKey = \sum publicKey_i * hash$

3.1.3) algorithm replace intermediate signatures with multisig and sharedPublicKey

3.1.4) algorithm save the record and update the root.

```
RecordModel {
 hash: '22839d5d994803eb15f8ffde5abbe62c2e0e2784efb53c837365ab94c08d8b55',
 timestamp: 1653722918550,
 timestampIndex: 0,
 key: '55874822b864c14',
 value: '55874822b864c14',
 version: 1,
 signaturesMap: {
  '03c73cf5d18b93bbd184a15a06d87044c2a39c80ca84835516c7203be942b03d9c' => '157c6c1c29bc20c20fcd4166824a78e255473a5a8d8c0c6d2162ffb61d54f400c'
 },
 signatureType: MULTISIG,
 publicKeys: [
  '022329c8cbd2d0784515a3bc457012a3b7ad7699b50bd906bcf999367e44a7b28b',
  '0363ce56b8ca32eb8c97db873649409e3bc5f6a452d2662e5a6f5e8d17acbf087b'
 ],
 stateHash: 'b8dabe0afae50be7975e73461172610e92016203d04554b3e10179705108495d'
}
```

Example of multisig record

### 4.3 Record validation

The validation process works as follows:

1) First signatures are validated: in case of intermediate signatures
    1.1) If signatures are intermediate, then for each intermediate signature the algorithm validate that: $publicKey_i * hash = signature * G$, where G – is a curve parameter (SECP256K1)
    1.2) If signature is multisig, then
        1.2.1) sharedPublicKey is reconstructed from involved public keys in signature process (the public keys with signatures are stored in record) and compared against received sharedPublicKey. If it's not equal – then validation is not passed
        1.2.2) then multisignature is validated as: $multiSignature * G = sharedPublicKey$

### 4.4 State consistency

To make sure, that all nodes have the same sets of records, the root has been introduced. The root is represented as sum of hashes of confirmed records (records with multisig): $root = \sum hash_i \bmod n$, where $n$ is a curve parameter. The following formula allows to build the root without order, so technically the append order of hashes doesn't make any sense in this case. Also, keep in mind, as algorithm has eventual consistency (without rollback option) – we can't guarantee any ordering.

Also, to make root update quick, the algorithm stores the root on record level:

1) On multisig record insert, the algorithm updates the root by addition of previous root to record's hash: $root = (root_{prev} + hash) \bmod n$
2) Then this root hash is appended to the record (I call it stateHash)
3) During next append of another new record, there is no need to recalculate the hash root of all records, but we sort confirmed (multisig) records in DESC order by timestamp and timestamp index, and take stateHash from the first record (which is the most recent one)

This approach is also useful for traceability and validation purpose, as all state can be replayed up to any point of history and calculated hash root can be compared with stateHash.

## 5. Replication process

The algorithm uses Gossip-like approach for replication and works as follows:

1) Gossiping happens in an infinite loop. On each loop (let's call it round) node (let's call it node A) choose randomly one peer (let's call it node B) with which it should gossip.
2) Then node A sends message to node B. The message includes timestamp and timestamp index from which node A expects to get updates from node B. It's like a pagination (i.e. give me all records, which timestamp is higher than <some timestamp>). Also keep in mind, that node A should store last requested timestamp and timestamp index for node B (in order not to request the same records again). If there wasn't communication before, then node A sends timestamp = 0 and timestampIndex = 0
3) Node B prepare an array of records which satisfy the provided timestamp and timestampIndex from node A (i.e. filter $record_B.timestamp >$

$record_A.timestamp \;||\; (record_A.timestamp === record_B.timestamp \;\&\&\; record_B.timestampIndex > record_A.timestampIndex))$

4) Node B sends records back to Node A as array of records with some limit (like no more than 10 records in one request, to prevent traffic flooding). The limit option can be set in algorithm parameters
5) Node A receive these records, sort them in ASC order, and apply them (the logic is described in "Append process / append from another node")
6) After append, node A updates last requested timestamp and timestampIndex for node B to the timestamp and timestampIndex from the last record received from node B (sorted in ASC order)
7) Then protocol awaits for random amount of milliseconds taken from range (the minimum and maximum time is specified in consensus parameters, check section 2). Once timeout expires – the process starts over

# 6. Proof of correctness

**6.1 Partial signature**

1) The general formula for obtaining partial signature is: $partialSignature = privateKey * hash$
2) The validation is: $partialSignatrue * G = publicKey * hash$
3) The correctness: $publicKey = privateKey * G => privateKey * hash * G = publicKey * hash => partialSignature * G = publicKey * hash$

**6.2 multi signature**
1) The general formula for building multisig is: $signature = \sum partialSignature_i$
2) The general formula for building multi public key is: $sharedPublicKey = \sum publicKey_i * hash$
3) The validation of signature: $signature * G = sharedPublicKey * hash$
4) The correctness of signature: $signature * G = sharedPublicKey * hash => \sum partialSignature_i * G = \sum publicKey_i * hash => partialSignature * G = publicKey * hash$

# 7. M-of-N connections

The following algorithm allows to build M-of-N connections network. This simply means, that there is no strict requirement, that all nodes should have connections between each other. This approach helps to define more efficient networking. However, keep in mind, that in order to guarantee consensus each node should have at least $f + 1$ connections.

# 8. Extensions and optimizations

## 8.1 Gossip to all peers

To speed up synchronization process, node may send messages to all known peers. This solution make sense when:

1) There are not so many nodes in the system (like 5-9)
2) The latency is predictable

## 8.2 Reducing Timestamp index

In case the solution use synchronization primitives and there is a guarantee that there won't be two or more records with the same timestamp, then timestampIndex may be reduced.

## 8.3 bitmap map for public keys

To reduce amount of traffic during replication, the algorithm uses bitmap as replacement for public keys. As all nodes should be aware of all public keys in network, it's fair to say, that all nodes have the same set of public keys. The bitmap algorithm (for the certain record's public key):

1) All public keys are sorted in ASC order
2) Then algorithm iterate over sorted public keys: in case the public key is present in record then algorithm return 1 otherwise 0. Example: there are public keys in network [A, B, C, D], the record includes signatures and public keys for [B, C], then bitmap will look: 0110 in binary form, or 6 in decimal form
3) This number in decimal is then used instead of public keys during replication process
4) The decoding happens in the opposite way